\documentclass{iopart} 

\usepackage{enumerate}

\usepackage[T1]{fontenc}

\expandafter\let\csname equation*\endcsname\relax
\expandafter\let\csname endequation*\endcsname\relax
\usepackage{amsmath, amsthm, amssymb, amsfonts}

\usepackage{graphicx}
\usepackage[colorlinks=true,linkcolor=blue,citecolor=blue,filecolor=cyan,urlcolor=blue,breaklinks=false]{hyperref}

\usepackage{cite}

\newcommand{\matr}[1]{\mathbf{#1}}
\newcommand{\timeorderedexp}[1]{\overrightarrow{\exp}\left({#1}\right)}

\newcommand{\period}{T}
\newcommand{\Affinity}{\mathcal{A}}

\newcommand{\ps}{p^\mathrm{ps}}
\newcommand{\js}{j^\mathrm{ps}_{ij}}
\newcommand{\jeff}{j^{\mathrm{eff}}_{ij}}
\newcommand{\peff}{p^\mathrm{eff}}

\newcommand{\vb}[1]{\textbf{#1}}
\newcommand{\order}[1]{\mathcal{O}(#1)}
\newcommand{\dd}{\mathrm{d}}
\newcommand{\vps}{\vb{p}^{\mathrm{ps}}}
\newcommand{\vpeff}{\vb{p}^\mathrm{eff}}

\newcommand{\occupation}{o_i}
\newcommand{\jump}{m_{ij}}
\newcommand{\djump}{\dot{m}_{ij}}

\newcommand{\etapd}{\hat{\eta}^\mathrm{ps}}
\newcommand{\etass}{\hat{\eta}^\mathrm{s}}

\newcommand{\ai}{\dot{a}_i}

\newcommand{\Pps}{\mathcal{P}}
\newcommand{\Ptilde}{\tilde{\mathcal{P}}}
\newcommand{\expval}[1]{\left\langle #1 \right\rangle}
\newcommand{\expvaltilde}[1]{\expval{#1}^\mathrm{aux}}

\newcommand{\ptilde}{\tilde{p}}
\newcommand{\vptilde}{\tilde{\vb{p}}}
\newcommand{\vjtilde}{\tilde{\vb{j}}}
\newcommand{\ktilde}{\tilde{k}_{ij}}
\newcommand{\jtilde}{\tilde{j}_{ij}}
\newcommand{\aptilde}{a^{\ptilde}_{ij}}
\newcommand{\jptilde}{j^{\ptilde}_{ij}}
\newcommand{\ts}{t^\mathrm{ps}_{ij}}
\newcommand{\ttilde}{\tilde{t}_{ij}}

\newcommand{\bound}{\sigma^\mathrm{eff}}
\newcommand{\boundtilde}{\tilde{\sigma}^\mathrm{eff}}
\newcommand{\arsinh}[1]{\mathrm{arsinh}\left(#1\right)}

\begin{document}
\title[]{A generalization of the thermodynamic uncertainty relation to
  periodically driven systems}
\author{Timur Koyuk$^1$, Udo Seifert$^1$, and Patrick Pietzonka$^{1,2}$}

\address{$^1$ II. Institut f\"ur Theoretische Physik, Universit\"at Stuttgart, 70550 Stuttgart, Germany}
\address{$^2$ DAMTP, Centre for Mathematical Sciences, University of Cambridge, Cambridge CB3 0WA, United Kingdom}

\begin{abstract}
  The thermodynamic uncertainty relation expresses a universal
  trade-off between precision and entropy production, which applies in
  its original formulation to current observables in steady-state
  systems. We generalize this relation to periodically time-dependent
  systems and, relatedly, to a larger class of inherently
  time-dependent current observables.  In the context of heat engines
  or molecular machines, our generalization applies not only to the
  work performed by constant driving forces, but also to the work
  performed while changing energy levels. The entropic term
  entering the generalized uncertainty relation is the sum of local
  rates of entropy production, which are modified by a factor that
  refers to an effective time-independent probability
  distribution. The conventional form of the thermodynamic uncertainty
  relation is recovered for a time-independently driven steady state
  and, additionally, in the limit of fast driving. We illustrate our
  results for a simple model of a heat engine with two energy levels.
  \\[1\baselineskip]
  \noindent{\it Dated\/}: \today
\end{abstract}
\vspace{1cm}

\paragraph{Introduction.} One of the main objectives of stochastic
thermodynamics is to relate thermodynamic properties of a small system
to the statistical fluctuations of its currents, for example the
mechanical work, dissipated heat or delivered chemical
output~\cite{jarz11,seif12}. A recent development in this spirit is
the thermodynamic uncertainty relation (TUR)
\begin{equation}
  \label{eq:TUR}
  D\sigma/J^2\geq k_\mathrm{B},
\end{equation}
which relates the relative fluctuations of a current, characterized by its diffusion
coefficient $D$ and average $J$, to the total rate of entropy
production $\sigma$~\cite{bara15,ging16}. In the following, we set Boltzmann's
constant $k_\mathrm{B}$ to unity. 

The TUR \eqref{eq:TUR} applies universally to current observables of steady-state
systems that can be modeled in continuous time using time-independent
Markovian dynamics, either on a discrete network or in continuous
space~\cite{pole16,nard17a,dech17}. While this covers already a large class of
systems and observables, recent efforts to push the limits of applicability of
the TUR even further have been fruitful, leading to variants for, e.g., finite
time~\cite{piet17,horo17} and first-passage time
fluctuations~\cite{garr17,ging17}.  However, there are various settings of stochastic
systems for which a direct application of the TUR fails, calling for
modifications that generalize Eq.~\eqref{eq:TUR}. Such settings include the
discrete-time case~\cite{shir17a,proe17,chiu18}, ballistic transport and
coherent dynamics~\cite{bran18}, and systems in linear response with
asymmetric Onsager matrices~\cite{maci18}.

In this paper, we focus on time-periodically driven systems as a
further prominent setting for which the conventional form
\eqref{eq:TUR} of the TUR does not hold~\cite{bara16}. Roughly
speaking, the driving protocol itself serves here as an exact external
clock that can enable the currents of the system proper to reach a
precision that surpasses the limit set by its rate of entropy
production. Hence, the TUR can be restored by adding the thermodynamic
cost for the external driving to the entropy production of the system
proper~\cite{bara17a}. Furthermore, systems driven by time-symmetric
protocols show similarities to the discrete-time case, allowing for a
generalization of the TUR in which the exponential of the entropy
production per period enters~\cite{proe17}. Recent work on large
deviation theory for arbitrary periodic driving has led to bounds on
the large deviation function for current
fluctuations~\cite{bert18,bara18a,bara18b}, which generalize similar
bounds that imply the TUR for time-independent driving~\cite{piet15}.

Applied to molecular motors and steady-state heat engines, the TUR yields a
fundamental bound on the efficiency, which depends only on the fluctuations of
measurable currents \cite{piet16b,piet17a}. However, paradigmatic models and
experimental realizations of stochastic heat engines often use
externally controlled, time periodic protocols~\cite{schm08,blic12}, to which
the TUR in its original formulation does not apply~\cite{holu18}. The
generalizations of the TUR following from the large deviation bounds in
Ref.~\cite{bara18b} apply to current observables that count jumps in
Markovian networks, which covers for example
the cycle current generated by a stochastic pump~\cite{raz16a,rots16}. Instead, the
current observables most relevant for heat engines are of a different
different type. In particular, the work performed on the system is given by
the change of the energy of the state that is currently occupied by the
system. The generalized thermodynamic uncertainty relation (GTUR) we derive
here applies to a broad class of current observables in periodically driven
systems, which includes the currents relevant for heat engines. In this
generalization, the entropy production $\sigma$ in Eq.~\eqref{eq:TUR} is
replaced by an effective entropy production, which can be larger than $\sigma$ and which
depends on a comparison between the currents in the periodic stationary
state and in a time-independent state of reference. We illustrate the GTUR and
an implied generalized bound on the efficiency for a simple two-level heat
engine that is alternatingly coupled to two different heat baths.

\paragraph{Setup.}
We consider a Markovian dynamics on a network of states with transition rates $k_{ij}(t) = k_{ij}(t+\period)$ from state $i$ to $j$ that are time-dependent and periodic with period $\period$.
These rates must be thermodynamically consistent and thus have to obey the local detailed balance condition
\begin{equation}
k_{ij}(t)/k_{ji}(t) = \exp(-\beta (t)\Delta_{ij}E(t)- \Affinity_{ij}(t)),
\label{eq:setup:local_detailed_balance_condition}
\end{equation}
where $\beta (t)$ is a possibly time-dependent inverse temperature,
$\Delta_{ij}E(t) \equiv E_j(t) - E_i(t)$ the energy difference between
internal states $i$ and $j$, and $\Affinity_{ij}(t)$ a driving affinity caused, e.g.,
by an external non-conservative force or a chemical reaction supplied by
chemostats.  These transition rates define a master equation
\begin{equation}
\dot{\matr{p}} (t) = \matr{L} (t) \matr{p} (t), \label{eq:setup:master_equation}
\end{equation} 
where the dot denotes a time-derivative and where the periodic matrix
$\matr{L} (t) = \matr{L} (t+\period)$ has the entries
\begin{align}
L_{ij} (t) \equiv k_{ji}(t) - \delta_{ij} r_i (t). \label{eq:setup:master_eq_matrix}
\end{align}
The entries $p_i(t)$ of vector $\vb{p}(t)$ in \eqref{eq:setup:master_equation}
give the probability that state $i$ is occupied at time $t$.  Furthermore,
$r_i(t) \equiv \sum_{j} k_{ij}(t)$ is the time-dependent exit rate and
$\delta_{ij}$ the Kronecker delta.  This periodically driven system converges
for long times into a periodic stationary state $\vps (t) = \vps (t+\period )$,
which is the unique periodic and normalized solution of
\eqref{eq:setup:master_equation}.

A stochastic trajectory $i(\tau)$ of length $t$ is characterized by an
occupation variable $\occupation (\tau)$, which is one if state $i$ is
occupied at time $\tau$ and zero, otherwise.  Note that we use a
notation that distinguishes the state $i$ from the trajectory
$i(\tau)$ by the argument. The variable $\jump (\tau)$ counts the directed
total number of jumps from $i$ to $j$ observed up to time $\tau$.  In
contrast to steady-state systems, a current can also depend on the
occupation $\occupation (\tau)$ and not only on jumps $\jump(\tau)$.
As an example, consider work that is performed while driving the
energy levels $E_i(\tau)$ without an external non-conservative force,
analogously to the definition of work used in the Jarzynski
relation~\cite{jarz97}.  The associated time-averaged power can be
expressed through the occupation variable as
\begin{equation}
P_i[i(\tau)] \equiv -\frac{1}{t} \int_0^t \dd{\tau}\occupation(\tau)\dot{E}_i(\tau),
\label{eq:setup:example_work_occ_traj}
\end{equation}
where we use the sign convention such that $P_i$ is positive when work
is delivered on average \textit{by} the system.  In the following,
such currents that only depend on the occupation are called
``occupation currents'', whereas currents that only depend on jumps
are called ``jump currents''.  An example for a jump current is the
entropy production~\cite{seif12}
\begin{equation}
\sigma[i(\tau)] \equiv \frac{1}{t}\int_0^t\dd{\tau}\sum_{i,j}\djump(\tau)\ln\frac{\ps_i(\tau)k_{ij}(\tau)}{\ps_j(\tau)k_{ji}(\tau)}.
\label{eq:setup:example_entropy_jump_traj}
\end{equation} 
A general current consisting of two parts, an occupation current and a jump current, reads
\begin{equation}
  j[i(\tau)] \equiv j_{\mathrm{occ}}[i(\tau)] + j_\mathrm{jump}[i(\tau)]\equiv \frac{1}{t}\int_0^t\dd{\tau} \sum_i \occupation (\tau) \ai(\tau)+ \frac{1}{t}\int_0^t\dd{\tau} \sum_{i,j}\djump (\tau) d_{ij}(\tau),
\label{eq:setup:time_add_curr}
\end{equation}
where $\ai(\tau)$ is the instantaneous change of a time-periodic
state variable $a_i(\tau)=a_i(\tau+T)$ and
$d_{ij}(\tau)=-d_{ji}(\tau) = d_{ij}(\tau + \period)$ is the increment
associated with a transition from $i$ to $j$ at time $\tau$.  Averages
of currents sampled over one period of the periodic stationary state
can be expressed as
\begin{equation}
J \equiv \expval{j[i(\tau)]} = \frac{1}{\period}\int_0^\period\dd{\tau}\sum_i \ps_i(\tau) \ai(\tau)+\frac{1}{\period}\int_0^\period\dd{\tau}\sum_{i>j} \js (\tau) d_{ij}(\tau), 
\label{eq:setup:stationary_current}%
\end{equation}
where $\expval{\cdot}$ denotes the average over all trajectories in
the periodic stationary state and
\begin{equation}
\js (\tau) \equiv \ps_i (\tau) k_{ij}(\tau) - \ps_j (\tau)k_{ji}(\tau)
\label{eq:setup:stationary_particle_current}
\end{equation}
denotes the periodic stationary probability current. We have used that
$\expval{\occupation (\tau)}=\ps_i (\tau)$ and
$\expval{\djump (\tau)} = \ps_i (\tau)k_{ij}(\tau)$.  The average of
the power~\eqref{eq:setup:example_work_occ_traj} delivered while the
system is in state $i$ is obtained for $\ai(\tau)=\dot{E}_i(\tau)$ and
reads
\begin{align}
P_i &\equiv \expval{P_i[i(\tau)]} = -\frac{1}{\period}\int_0^\period\dd{\tau} \ps_i (\tau) \dot{E}_i(\tau).\label{eq:setup:definition_power}
\end{align}
The average of the fluctuating entropy production in
\eqref{eq:setup:example_entropy_jump_traj} is obtained for
$d_{ij}(\tau)=\ln[{\ps_i(\tau)k_{ij}(\tau)}/{\ps_j(\tau)k_{ji}(\tau)}]$, yielding the
average rate of entropy production
\begin{equation}
\sigma\equiv\expval{\sigma[i(\tau)]} = \frac{1}{\period}\int_0^\period\dd{\tau} \sum_{i>j} \js (\tau)\ln\frac{\ps_i(\tau)k_{ij}(\tau)}{\ps_j(\tau)k_{ji}(\tau)}.\label{eq:setup:definition_entropy}
\end{equation}

Fluctuations of currents in the ensemble of trajectories $i(\tau)$
with $0\leq \tau\leq t$  are quantified via the scaled cumulant generating function
\begin{equation}
\lambda_t(z)\equiv\frac{1}{t}\ln\expval{e^{ztj[i(\tau)]}},
\label{eq:setup:scgf}
\end{equation}
which is in the following referred to as the ``generating
function''. Its long-time limit is
$\lambda(z) \equiv \lim\limits_{t\to\infty}\lambda_t(z)$.  Denoting
derivatives for $z$ with $'$, the average current follows as
$J=\lambda'(0)$ and the diffusion coefficient associated with that
current is given by
\begin{equation}
  D\equiv\lim_{t\to\infty}t \expval{\left(j[i(\tau)]-\expval{j[i(\tau)]}\right)^2}/2=\lambda''(0)/2.
  \label{eq:setup:diffusion_coefficient}
\end{equation}
The calculation of these quantities using time-ordered exponentials is
sketched in~\ref{sec:appa}.

\paragraph{Main result.}
Our main result generalizes the TUR \eqref{eq:TUR} to systems driven into a
periodic stationary state and is called in the following the generalized
thermodynamic uncertainty relation (GTUR).  It is valid for all currents
defined in \eqref{eq:setup:time_add_curr}.  The GTUR reads
\begin{equation}
D \bound /J^2\ge 1 \label{eq:main_result:bound}
\end{equation}
with the effective rate of entropy production
\begin{align}
&\bound \equiv \frac{1}{\period}\int_0^\period\dd{\tau}\sum_{i>j} \left(\frac{\jeff (\tau)}{\js (\tau)}\right)^2 \sigma^\mathrm{ps}_{ij}(\tau).\label{eq:main_result:definition_of_bound}
\end{align}
Here,
\begin{align}
\sigma^\mathrm{ps}_{ij}(\tau) &= \js (\tau)\ln\frac{\ps_i(\tau)k_{ij}(\tau)}{\ps_j (\tau) k_{ji}(\tau)} \label{eq:main_result_sigma_ij}
\end{align}
is the instantaneous periodic stationary entropy production rate associated with the
link $ij$.  The term $\jeff(\tau)$ is an effective current
\begin{align}
\jeff (\tau) &\equiv \peff_i k_{ij}(\tau) - \peff_j k_{ji}(\tau) \label{eq:main_result:jeff}
\end{align}
caused by a time-independent effective density $\peff_i$. It will in
general not satisfy a conservation law.  The effective density is a
set of free variation parameters that have to fulfill the condition
$\sum_i \peff_i = 1$.  For time-independent transition rates, the
effective densities can be chosen as the stationary state
$p^\mathrm{s}_i$.  Then, the effective currents $\jeff(\tau)$ are the
stationary ones and $\bound=\sigma$.  Hence,
\eqref{eq:main_result:bound} assumes the conventional form of the
TUR. 

For an exact experimental determination of $\bound$, it
is necessary to extract all phase-dependent probabilities and currents
from a long trajectory that spans many periods. Nonetheless, a lower
bound on $\bound$ can be obtained from the measurement of the
diffusivity $D$ and average $J$ of any accessible current of the system.

We emphasize that the bound~\eqref{eq:main_result:bound} has a broader
applicability than two earlier generalizations of the TUR. First, it
is not restricted to time-symmetric driving as the one in
Ref.~\cite{proe17}. Second, our generalization applies not only to
currents with time-independent increments, which Ref.~\cite{bara18b}
focuses on.  Consequently, as we will show below, our bound on
precision is non-trivial for two-level systems, where all currents
with time-independent increments must vanish. Interestingly, for
time-dependent increments, Ref.~\cite{bara18b} provides a variant of
the TUR that replaces not only the entropy production by a modified
one but also the average current $J$. However, this modified current is liable to become zero
for the most relevant currents in heat engines.

Two different choices for $\vb{p}^\mathrm{eff}$ have an immediate physical interpretation.
The first choice is defined through
\begin{align}
\left(\frac{1}{\period}\int_0^\period\dd{\tau}\matr{L} (\tau)\right) \vb{p}^\mathrm{eff} &= 0 \label{eq:main_result:p_eff_a}
\end{align}
as the stationary solution of the master equation with
time-averaged transition rates.  
If the driving frequency $\omega\equiv 2\pi/T$ is large compared to the entries of~$\matr{L}(\tau)$,
the periodic stationary state $\vps(\tau)$ converges to this effective
density $\vpeff$, see~\ref{sec:appb}. 

The second choice for the variation parameters $\vpeff$ is a simple time average over the periodic stationary state
\begin{align}
\vpeff &= \frac{1}{\period}\int_0^\period\dd{\tau}\vps (\tau),\label{eq:main_result:p_eff_b}
\end{align}
i.e., the average fraction of the total time spent in a state $i$ during one period.

For these two choices, the corresponding relation
\eqref{eq:main_result:bound} can be regarded as a genuine
generalization of the conventional TUR, which is restored for
time-independent transition rates, where
$\vb{p}^\mathrm{eff}=\vb{p}^\mathrm{s}$ holds by
construction. Physically, the effective entropy production
\eqref{eq:main_result:definition_of_bound} may be interpreted as a
modification of the actual entropy production
\eqref{eq:setup:definition_entropy}. This modification is mediated by
the term $({\jeff (\tau)}/{\js (\tau)})^2$, which encodes the
``distance'' from a system in a time-independent state. In particular,
for zero or small affinities $\mathcal{A}_{ij}$ the tendency of the
system to relax towards an instantaneous stationary state reduces the
absolute value of $\js (\tau)$ with respect to that of $\jeff (\tau)$
for most times $\tau$ and links $ij$, such that $\bound>\sigma$ holds
for the vast majority of possible driving protocols. Other choices for
$\vpeff$, \textit{e.g.}, a uniform distribution, a delta distribution,
or even a choice where some of the $\peff_i$ are negative are
conceivable. However, such choices do generally not yield the
conventional TUR for time-independent rates and therefore lack the
interpretation of $\bound$ being different from $\sigma$ as an
indicator for time-dependence.

The two choices~\eqref{eq:main_result:p_eff_a}
and~\eqref{eq:main_result:p_eff_b} become equivalent in the limiting
case of large driving frequencies $\omega$ or for linear response
around a genuine non-equilibrium steady state.  In leading order, as
discussed in~\ref{sec:appb}, the periodic stationary state
$\vps (\tau)$ is then time-independent and solves
Eq. \eqref{eq:main_result:p_eff_a}. Consequently, the currents
$\js(\tau)$ and $\jeff(\tau)$ become the same, which leads to
$\bound=\sigma$ and thus restores the original form \eqref{eq:TUR} of
the TUR. However, in those limiting cases where both currents vanish
in zeroth order, in particular in linear response around an
equilibrium state, $\js(\tau)$ and $\jeff(\tau)$ differ in leading
order and $\bound$ remains different from $\sigma$.

\begin{figure}[tbp]
\centering\includegraphics[width=0.75\textwidth]{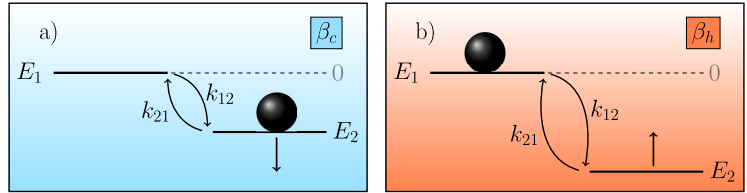}
\caption{ Schematic representation of the two-level heat engine: (a) In the first half of
  the period $E_2(t)$ decreases from its initial value $\epsilon_0+E$ at fixed
  cold inverse temperature $\beta_c$, whereas in the second half, (b), it
  increases from $\epsilon_0-E$ at hot inverse temperature $\beta_h$.  }
\label{fig:ill_and_appl:scheme_particle}
\end{figure}

\paragraph{Illustration: Two level heat engine.}
We consider a heat engine that is coupled alternatingly to two different heat baths.
It has two states with one energy periodically driven, such that
\begin{equation}
E_1(t) = 0\qquad \textup{and}\qquad E_2(t) = E\cos(\omega t) + \epsilon_0.
\end{equation}
Here, $E$ is an amplitude and $\epsilon_0$ an offset with respect to
the energy of the first state.  In the first half of the period,
$0\leq t< T/2$, the temperature $\beta(t)$ is fixed at a cold
inverse temperature $\beta_c$ and in the second half,
$T/2\leq t< T$, it is fixed at a hot inverse temperature $\beta_h<\beta_c$.  We
choose the individual rates symmetrically according to the local
detailed balance condition in
\eqref{eq:setup:local_detailed_balance_condition} as
\begin{equation}
k_{ij}(t) = k_0 \exp(-\beta (t)\Delta_{ij}E(t)/2),
\label{eq:ill_and_appl:rates}
\end{equation}
where $k_0$ determines the basic time scale for particle jumps.
A schematic representation of the engine is shown in fig.~\ref{fig:ill_and_appl:scheme_particle}.

\begin{figure}[tbp]
\centering
\includegraphics[width=0.8\textwidth]{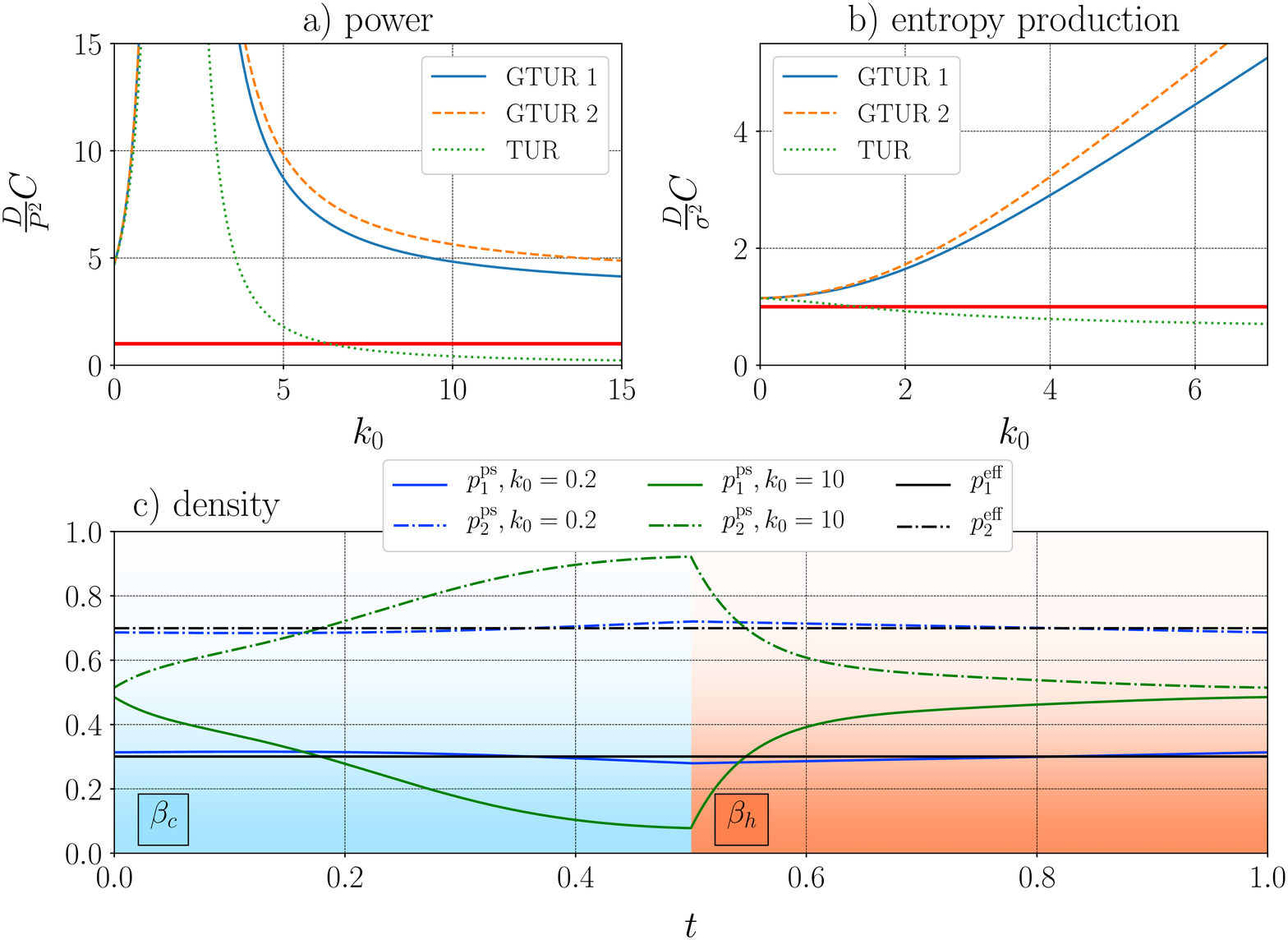}
\caption{Two-level heat engine with $E=1.0$, $\epsilon_0=-1.5$,
  $\beta_h=0.1$, $\beta_c=1$, and $T=1$. Generalized uncertainty
  relations with $C=\bound$ are shown as a function of rate amplitude
  $k_0$ for the two different choices
  in~\eqref{eq:main_result:p_eff_a} (GTUR~1)
  and~\eqref{eq:main_result:p_eff_b} (GTUR~2) compared to the TUR with
  $C=\sigma$. The value 1 on the right hand sides of the relations is
  marked by a solid red line. Current of interest in panel (a) is the
  power, for which the TUR is violated for $k_0 \gtrsim 6.0$. A
  singularity occurs at $k_0\simeq 1.75$, where the power passes
  zero. Panel (b) corresponds to the entropy production, for which the
  TUR is violated for $k_0 \gtrsim 1.8$. The periodic stationary state for
  two selected values of $k_0=0.2$ and $10$ is shown in panel (c),
  revealing kinks when the inverse temperature switches
  between~$\beta_c$ and~$\beta_h$. The effective density according to
  the choice~\eqref{eq:main_result:p_eff_a} is approached for small
  rates $k_0$.}
\label{fig:ill_and_appl:gtur_k0}
\end{figure}

For the analysis shown in Fig.~\ref{fig:ill_and_appl:gtur_k0}, we vary
the rate amplitude $k_0$ and keep all other parameters fixed. The
periodic stationary distribution yielding $\sigma$, $\bound$, and $P$
and the diffusion constants $D$ for the respective currents are
calculated numerically using the methods outlined
in~\ref{sec:appa}. The left-hand side
(l.h.s) of the GTUR~\eqref{eq:main_result:bound} for the two choices
in \eqref{eq:main_result:p_eff_a} and \eqref{eq:main_result:p_eff_b}
as well as the l.h.s. of the corresponding steady state
TUR~\eqref{eq:TUR} are shown for the power (Fig.~\ref{fig:ill_and_appl:gtur_k0}a) and for the entropy
production (Fig.~\ref{fig:ill_and_appl:gtur_k0}b) as currents of interest.

For small $k_0$, i.e., in the fast driving limit $k_{ij}(t)\ll\omega$,
the periodic stationary state approaches a time-independent state, as
shown in Fig.~\ref{fig:ill_and_appl:gtur_k0}c.  Then, the two choices
for the GTUR and the TUR become identical for small $k_0$, as explained
in~\ref{sec:appb}.  Differences between the two choices for $\vpeff$
can be seen for larger $k_0$.  In this regime, the choice
\eqref{eq:main_result:p_eff_a} becomes better than the choice
\eqref{eq:main_result:p_eff_b} for both currents.  Furthermore, the
TUR for power is strongly violated for large $k_0$.  Here, the GTUR
does hold and becomes sharper again.  For the entropy production, the
GTUR is less sharp for large $k_0$ where again the TUR does not hold.

\paragraph{Bound on efficiency of heat engines.}
The trade-off relation between power, efficiency and constancy,
derived in \cite{piet17a} as a consequence of the TUR, applies to
steady-state heat engines, but in general not to periodically driven
systems~\cite{holu18}.  The GTUR derived here generalizes this
trade-off relation and bounds the efficiency of periodically driven
heat engines as we show in the following. The formally similar
trade-off described in Ref.~\cite{shir16} applies to periodically
driven engines, but does not make reference to power fluctuations.

The efficiency of a heat engine is given by
\begin{align}
\eta \equiv P/\dot{Q}_\mathrm{in} \le \eta_C \equiv 1 - \beta_h/\beta_c,
\label{eq:ill_and_appl:eta_eta_C}
\end{align}
where $P\equiv\sum_iP_i$ is the total output power of the heat engine
defined in \eqref{eq:setup:definition_power} and $\dot{Q}_\mathrm{in}$ is the
heat current flowing into the system from the hot reservoir.  This efficiency is
always bounded by the Carnot efficiency $\eta_C$.  Following the analogous
calculations from Ref.~\cite{piet17a}, the efficiency of a periodically driven
heat engine $\eta$ is bounded due to the GTUR~\eqref{eq:main_result:bound} by
the stronger relation
\begin{equation}
\eta \le \etapd\equiv \frac{\eta_C}{1 + P\sigma/\left(\beta_c D_P\bound\right)} \le \eta_C,
\label{eq:ill_and_appl:bound_on_eta}
\end{equation}
where $D_P$ is the diffusion coefficient~\eqref{eq:setup:diffusion_coefficient}
of the fluctuating output power.  

\begin{figure}[tbp]
\centering
\includegraphics[width=0.8\textwidth]{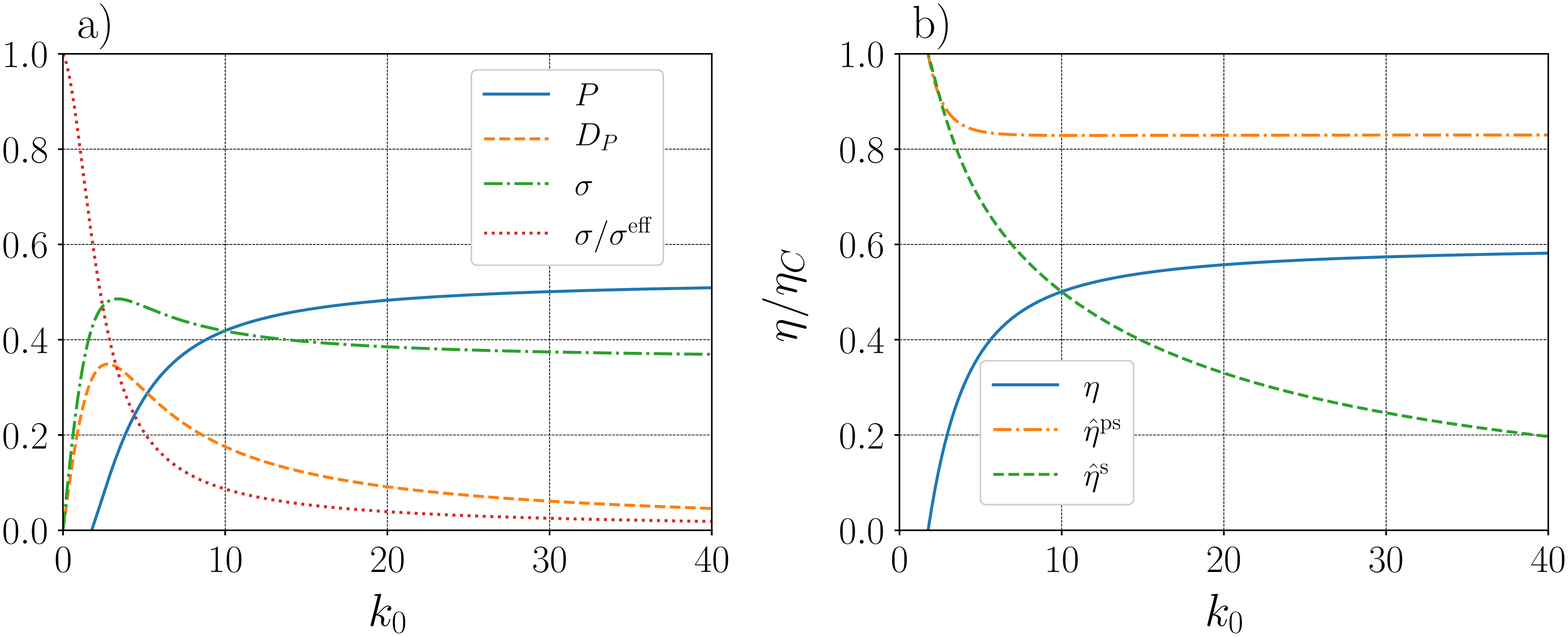}
\caption{Further data for the two-level heat engine. (a) Power $P$,
  its diffusion coefficient $D_{P}$, the actual entropy production
  $\sigma$ and its ratio to the effective one~$\bound$ as a function
  of the rate amplitude $k_0$, (b) efficiency $\eta$, bound on
  efficiency for periodically driven systems $\etapd$, and the
  steady-state heat engine bound $\etass$ as functions $k_0$.  The
  engine produces a positive output power for $k_0\gtrsim 1.75$. For
  rate amplitudes $k_0\gtrsim 10$ the steady-state heat engine bound
  does not hold.  The energy amplitude $E=1.0$, the offset energy
  $\epsilon_0=-2.0$, and the temperatures $\beta_c=1$ and
  $\beta_h=0.1$ are fixed.  }
\label{fig:ill_and_appl:eta_heat_engine}
\end{figure}
As an example, we consider the heat engine from
fig.~\ref{fig:ill_and_appl:scheme_particle} and vary the rate
amplitude~$k_0$. The effective entropy production $\bound$ is
calculated from the choice~\eqref{eq:main_result:p_eff_a} for the
effective density.  The quantities entering the
bound~\eqref{eq:ill_and_appl:bound_on_eta} are shown in
fig.~\ref{fig:ill_and_appl:eta_heat_engine}a and the efficiency $\eta$
of the heat engine and the bound $\etapd$ are shown in
fig.~\ref{fig:ill_and_appl:eta_heat_engine}b. This bound is compared
to the naive bound valid for steady-state engines, called here
$\etass$ and defined just as $\etapd$, but with $\bound$ replaced by
the true entropy production $\sigma$.  For small rate amplitudes
$k_0\ll\omega$, where the GTUR assumes the form of the TUR, the new
bound $\etapd$ based on the GTUR becomes identical to $\etass$.  In the
regime of large rate amplitudes, $k_0\gg \omega$, both the efficiency
and the power increase to finite limiting values, see \ref{sec:appb}. Since at the same
time fluctuations decrease, the bound $\etapd$ becomes rather strong
with the actual efficiency being only about $25\,\%$ below this bound,
whereas the bound $\etass$ no longer holds.

\paragraph{Derivation.}
We now derive our main result shown in \eqref{eq:main_result:bound}.  For this purpose,
we bound the generating function by introducing an auxiliary dynamics with
path weight $\Ptilde[i(\tau)]$.  A similar formalism has been
introduced in Ref.~\cite{dech17} for continuous degrees of freedoms.  The
weight of paths from the periodic stationary state is denoted by
$\Pps[i(\tau)]$, so that
\begin{equation}
\ps_i(t) =\expval{o_i(t)}= \sum_{i(\tau)} \Pps[i(\tau)]\occupation(t),\quad
\ps_i(t)k_{ij}(t) = \expval{\djump(t)}=\sum_{i(\tau)}\Pps[i(\tau)]\djump(t),\label{eq:derivation:stationary_path_weight}
\end{equation}
where the summation indicates a path integral over all trajectories
$i(\tau)$, and where the occupation variable $o_i(t)$ and the jump
variable $m_{ij}(t)$ refer implicitly to these trajectories.  We split
up $\Pps[i(\tau)]$ as $\Pps[i(\tau)] = \Pps[i(\tau)|i_0]\ps_{i_0}(0)$,
where $\Pps[i(\tau)|i_0]$ is the path weight conditioned in the system
being in state $i_0$ at time $t=0$, which in turn is associated with
the probability $\ps_{i_0}(0)$.  Likewise, for the path weight of the
auxiliary dynamics, we split
$\Ptilde[i(\tau)] = \Ptilde[i(\tau)|i_0]\ptilde_{i_0}(0)$, with an
\textit{a priori} arbitrary initial distribution $\ptilde_{i_0}(0)$.

The generating function in \eqref{eq:setup:scgf} for a current $j[i(\tau)]$ can be written in terms of $\Pps[i(\tau)]$ and $\Ptilde[i(\tau)]$ as
\begin{align}
\lambda_t(z) &= \frac{1}{t}\ln\expval{e^{ztj[i(\tau)]}}= \frac{1}{t}\ln\expvaltilde{\frac{\Pps[i(\tau)]}{\Ptilde[i(\tau)]}e^{ztj[i(\tau)]}}\notag\\
&=\frac{1}{t}\ln\expvaltilde{\exp\left(ztj[i(\tau)] - \ln\frac{\Ptilde[i(\tau)|i_0]}{\Pps[i(\tau)|i_0]} -\ln\frac{\ptilde_{i_0}(0)}{\ps_{i_0}(0)}\right)}, \label{eq:derivation:lambda_tilde_dynamic}
\end{align}
where $\expvaltilde{\cdot}$ denotes the average over all trajectories in the auxiliary dynamics.
This generating function can be bounded by using Jensen's inequality as
\begin{align}
\lambda_t(z) \ge &z\expvaltilde{j[i(\tau)]}-\frac{1}{t}\expvaltilde{\ln\frac{\ptilde_{i_0}(0)}{\ps_{i_0}(0)}}
-\frac{1}{t}\expvaltilde{\ln\frac{\Ptilde[i(\tau)|i_0]}{\Pps[i(\tau)|i_0]}}.
\label{eq:derivation:bound_lamdba_jensen}
\end{align}

Next, the path weight $\Ptilde[i(\tau)]$ is chosen such that it produces in
analogy to \eqref{eq:derivation:stationary_path_weight} a density $\vptilde(t)$,
which is the periodic stationary solution for the auxiliary rates
$\ktilde(t)$, i.e.,
\begin{equation}
  \ptilde_i(t) =\expvaltilde{o_i(t)}= \sum_{i(\tau)} \Ptilde[i(\tau)]\occupation(t),\quad
  \ptilde_i(t)\tilde{k}_{ij}(t) =
  \expvaltilde{\djump(t)}=\sum_{i(\tau)}\Ptilde[i(\tau)]\djump(t).
  \label{eq:derivation:stationary_path_weight_tilde}
\end{equation}
In the following, we denote the currents associated with \eqref{eq:derivation:stationary_path_weight_tilde} as 
\begin{equation}
\jtilde(t) \equiv \ptilde_i(t)\tilde{k}_{ij}(t)-\ptilde_j(t)\tilde{k}_{ji}(t)
\label{eq:derivation:jtilde}
\end{equation}
 and the corresponding traffic, or activity, as 
\begin{equation}
\ttilde(t)\equiv \ptilde_i(t)\tilde{k}_{ij}(t)+\ptilde_j(t)\tilde{k}_{ji}(t).
\label{eq:derivation:ttilde}
\end{equation} 

The first term in Eq.~\eqref{eq:derivation:bound_lamdba_jensen}, i.e.,
the
current~\eqref{eq:setup:time_add_curr} averaged by the auxiliary dynamics,
then reads
\begin{equation}
\expvaltilde{j[i(\tau)]} = \int_0^t\dd{\tau} \sum_{i>j}\jtilde(\tau) d_{ij}(\tau) + \sum_i \ptilde_i(\tau) \ai(\tau).
\label{eq:derivation:aux_current} 
\end{equation}
The second term in \eqref{eq:derivation:bound_lamdba_jensen} can be written as
a Kullback-Leibler divergence between the initial distribution $\vps(0)$ of the
original periodic stationary distribution and the initial distribution
$\vptilde(0)$ of the auxiliary dynamics
\begin{align}
D\left(\vptilde(0)\vert\vert\vps(0)\right) &\equiv \sum_i \ptilde_i(0)\ln\frac{\ptilde_i(0)}{\ps_i(0)} \ge 0. \label{eq:derivation:kullback_leibler_divergence}
\end{align}
We evaluate the third term in \eqref{eq:derivation:bound_lamdba_jensen} by
calculating the fraction of the two path weights for the same trajectory
$i(\tau)$
\begin{align}
\frac{\Ptilde[i(\tau)|i_0]}{\Pps[i(\tau)|i_0]} =& \exp\left( \int_0^t\dd{\tau} \sum_{ij} \djump(\tau)\ln\frac{\widetilde{k}_{ij}(\tau)}{k_{ij}(\tau)} - \sum_i \occupation(\tau)\left(\tilde{r}_i(\tau)-r_i(\tau)\right) \right),
\label{eq:derivation:fraction_of_path_weights}
\end{align}
where $\tilde{r}_i(\tau)\equiv \sum_j\widetilde{k}_{ij}(\tau)$ are the exit
rates of the auxiliary dynamics.  Inserting
\eqref{eq:derivation:fraction_of_path_weights} into
\eqref{eq:derivation:bound_lamdba_jensen} leads to terms containing averages
with the path weight $\Ptilde[i(\tau)]$ for $\occupation(\tau)$ and
$\djump(\tau)$, for which we can use
\eqref{eq:derivation:stationary_path_weight_tilde}.  We express the
rates in terms of the current and the associated traffic as
\begin{align}
\tilde{k}_{ij}(t) = \left(\jtilde(t) + \ttilde(t)\right)/(2\ptilde_i(t)).
\label{eq:derivation:ktilde_current_traffic}
\end{align}
After optimizing the third term in \eqref{eq:derivation:bound_lamdba_jensen}
with respect to the traffic, the optimal rates read
\begin{align}
\ktilde^*(t) = \left(\jtilde(t)+\sqrt{\left(\jtilde(t)\right)^2+4\ptilde_i(t)\ptilde_j(t)k_{ij}(t)k_{ji}(t)}\right)/(2\ptilde_i(t)).
\label{eq:derivation:choice_of_rates}
\end{align}
Finally, using these auxiliary rates and inserting \eqref{eq:derivation:choice_of_rates} into \eqref{eq:derivation:bound_lamdba_jensen} leads to a bound in terms of $\vjtilde(t) \equiv\{\jtilde (t)\}$ and $\vptilde(t)\equiv\{\ptilde_i(t)\}$,
\begin{align}
  \lambda_t(z)
  \ge&z\expvaltilde{j[i(\tau)]}-\frac{1}{t}\int_0^t\dd{\tau}L\left(\vptilde(\tau),\vjtilde
    (\tau)\right) -\frac{1}{t}D\left(\vptilde(0)\vert\vert\vps(0)\right),
\label{eq:derivation:large_dev_analogue}
\end{align}
with 
\begin{align}
  L\left(\vptilde(\tau),\vjtilde(\tau)\right) \equiv\sum_{i>j}&\, \jtilde(\tau) \left(\arsinh{\frac{\jtilde(\tau)}{\aptilde(\tau)}}-\arsinh{\frac{\jptilde(\tau)}{\aptilde(\tau)}}\right)\notag\\
  &-\left(\sqrt{(\aptilde(\tau))^2 + (\jtilde(\tau))^2} -
    \sqrt{(\aptilde(\tau))^2 +
      (\jptilde(\tau))^2}\right), \label{eq:derivation:large_dev_L}
\end{align}
where
\begin{equation}
\jptilde(\tau)\equiv \ptilde_i(\tau) k_{ij}(\tau)-\ptilde_j(\tau)
k_{ji}(\tau), \ 
\aptilde(\tau)\equiv \sqrt{4\ptilde_i(\tau)\ptilde_j(\tau)k_{ij}(\tau)k_{ji}(\tau)}.
\end{equation}
The densities $\vptilde(t)$ and currents $\vjtilde(t)$ of the auxiliary dynamics must fulfill the conditions
\begin{align}
\sum_i \ptilde_i(t) = 1, \quad \ptilde_i(t) > 0 \quad \mathrm{and}\quad
\dot{\ptilde}_i(t) = -\sum_j \jtilde(t)
\label{eq:derivation:condition_aux_dyn} 
\end{align}
for all $i$ and $t$, which guarantee that a matching set of auxiliary
transition rates $\tilde{k}_{ij}(t)$ can be found.

Now, we choose a suitable ansatz for the densities $\vptilde(t)$ and currents $\vjtilde(t)$ of the auxiliary dynamics.
One can easily verify that the ansatz
\begin{equation}
\ptilde_i(t) = \ps_i(t) + \epsilon \left(\ps_i(t)-\peff_i\right),\qquad
\jtilde(t) = \js(t) + \epsilon \js(t)
\label{eq:derivation:ansatz}
\end{equation}
with an arbitrary small optimization parameter $\epsilon =\order{z}$ fulfills the
conditions \eqref{eq:derivation:condition_aux_dyn}, if $\sum_i\peff_i=1$.
Using this ansatz, one can expand \eqref{eq:derivation:large_dev_analogue} up
to order $\order{z^2}$ for small $z$ to obtain a local bound on the generating
function after an optimization with respect to the parameter
$\epsilon$.  Additionally, we restrict ourselves to observation times $t=nT$ that are
multiples of the period.  Then, \eqref{eq:derivation:large_dev_analogue}
reads up to $\order{z^2}$
\begin{align}
\lambda_{n\period}(z) &\ge z\left(J + z\frac{J^2}{2\boundtilde(n\period)}\right) + \order{z^3},\notag\\
\boundtilde(n\period) &\equiv \frac{1}{\period} \int_0^{\period}\dd{\tau}\sum_{i>j}\left(\frac{\jeff(\tau)^2}{\ts(\tau)}\right) +\frac{1}{n\period}\sum_i\frac{\left(\ps_i(0)-\peff_i\right)^2}{\ps_i(0)},
\label{eq:derivation:bound_on_lamdba}
\end{align}
where $\ts(\tau)$ is the stationary traffic and $J$ the stationary current
\eqref{eq:setup:stationary_current}.  This is our strongest and most general
result, holding for finite time after $n$ periods, small values of $z\sim 0$
and currents defined in \eqref{eq:setup:time_add_curr}.

Using $D_{n\period} \equiv \lambda''_{n\period}(0)/2$ as a
finite-time generalization of the diffusion
coefficient~\eqref{eq:setup:diffusion_coefficient}, the local
quadratic bound in \eqref{eq:derivation:bound_on_lamdba} implies an
inequality on precision for an arbitrary current as
\begin{equation}
2D_{n\period} \boundtilde(n\period)/J^2 \ge 1.
\label{eq:derivation:percision_inequality}
\end{equation}
Using the inequality $\ts(\tau) \leq 2\js(\tau)^2/\sigma^\mathrm{ps}_{ij}(\tau)$,
one obtains the bound given in \eqref{eq:main_result:bound} with an
additional term arising from the Kullback-Leibler divergence. Taking
the long-time limit $n\to\infty$, one obtains exactly the GTUR in
Eq.~\eqref{eq:main_result:bound} with the effective entropy
production~\eqref{eq:main_result:definition_of_bound}.

As an aside, we note that in the case of time-independent rates the
ansatz~\eqref{eq:derivation:ansatz} becomes
$\ptilde_i = p_i^\mathrm{s}$,
$\jtilde = (1+\epsilon) j^\mathrm{s}_{ij}$, where the upper index
``s'' denotes the stationary distribution and currents.  Using a
quadratic bound \cite{ging16,ging16a} on
$L(\vptilde(\tau),\vjtilde(\tau))$ in
\eqref{eq:derivation:large_dev_L}, and performing an optimization with
respect to $\epsilon$ leads to the quadratic bound on $\lambda_t(z)$,
which implies the finite-time TUR \cite{piet17,horo17}.  In the
long-time limit $t\to\infty$, this lower bound on the generating
function $\lambda(z)$ becomes equivalent to the upper bound on the
large deviation function \cite{piet15,ging16}. In Ref.~\cite{bara18b},
such a quadratic bound on $L(\vptilde(\tau),\vjtilde(\tau))$ has led
to a global bound on the large deviation function for jump currents in
a periodically driven system. The corresponding local bound, though
formally similar, is different from the GTUR derived here.

Unlike most variants of the TUR, the present
generalization~\eqref{eq:main_result:bound} is not a consequence of a
simple, usually quadratic, global bound on the large deviation
function or generating function. Technically, choosing small $z$ and
consequently small $\epsilon$ in
Eq.~\eqref{eq:derivation:bound_on_lamdba} is necessary to ensure that
the specific ansatz \eqref{eq:derivation:ansatz} for the density is
positive. From a more general perspective, we note that the
fluctuations of occupation currents are always limited to a finite
range that is set by those realizations of $o_i(\tau)$ that maximize
or minimize $j_\mathrm{occ}[i(\tau)]$ in Eq.~\eqref{eq:setup:time_add_curr}, which rules out the
existence of any global quadratic upper bound on the large deviation
function.

Finally, the local quadratic bound in
\eqref{eq:derivation:bound_on_lamdba} is valid for small enough $z$
and also at finite time.  This leads to
Eq.~\eqref{eq:derivation:percision_inequality} as a generalization of
the finite-time uncertainty relation \cite{piet17,horo17} to
periodically driven systems.  Here, the Kullback-Leibler divergence
\eqref{eq:derivation:kullback_leibler_divergence}, which leads to the
second term of $\boundtilde\left(n\period\right)$, does not vanish.
This term quantifies the difference between the initial distribution
of the periodic stationary state and an effective time-independent
distribution.  For large driving frequencies the periodic stationary
state $\vps(t)$ converges to an effective density $\vpeff$, as we show
in~\ref{sec:appb}, and hence the Kullback-Leibler divergence
vanishes. Moreover, the Kullback-Leibler divergence can be brought to
vanish by choosing $\vpeff=\vps(0)$.

\paragraph{Conclusion.}
We have generalized the TUR to time-periodically driven systems and to
a larger class of current observables. This class includes the
currents most relevant for periodically driven heat engines,
in particular the work associated with changing the energy level
of a state occupied by the system. 

Our generalization restores the ordinary form of the TUR for the special case of large driving
frequencies.  Hence, for large
driving frequencies precision has a universal minimal cost.  This is
somewhat remarkable, because although the system can be described by a
time-independent distribution a one-to-one mapping of a periodically driven system
to a steady-state system fails at the description of currents.
Thus, we extend the applicability of the TUR and the ensuing trade-off
relations to heat engines driven solely by fast alterations of energy
levels and temperature.

For moderate or low driving frequencies one has to compare the
periodic stationary currents with the associated effective currents in
a time-independent state of reference. One can then predict whether a
larger entropy production or smaller currents than the effective ones
are needed for a higher precision.  Furthermore, due to the
generalization of the TUR, one can bound the efficiency of heat
engines and hence predict whether an engine is able to work close to
Carnot efficiency or not.  Finally, we note that in a setting where
$\beta_c=\beta_h=1$, the bound on efficiency for heat engines can be
adapted to isothermal engines transforming work to heat or work to
work.

A formulation of the GTUR for overdamped Brownian motion is straightforward, either by performing the
continuum limit on a finely discretized state space or by redoing the
derivation using the path weights pertaining to Langevin dynamics.

\paragraph{Acknowledgments.} Work funded in part by the ERC
under the EU Horizon 2020 Programme via ERC grant agreement 740269.

\begin{appendix}
  \section{Calculation of cumulants} 
  \label{sec:appa}
  The solution of the time-dependent master
  equation~\eqref{eq:setup:master_equation} for an initial
  distribution $\vb{p}(0)$ is formally given by the
  time-ordered exponential
  \begin{align}
    \vb{p} (t) &= \timeorderedexp{\int_0^t \dd{\tau}\matr{L} (\tau)} \vb{p}(0)\equiv\vb{M} (t) \vb{p}(0) \label{eq:main_result:time_ordered_exp},
  \end{align}
  defining the evolution operator $\vb{M}(t)$.
  There exists a unique initial condition $\vb{p}^\mathrm{ps}(0)$ that
  corresponds to the periodic stationary state $\vps (t)$.  This
  initial condition can be determined using the periodicity of
  $\vps (t)$, which leads to the eigenvalue equation
  \begin{equation}
    \vb{p}^\mathrm{ps}(0)=\vb{M}(\period)\vb{p}^\mathrm{ps}(0). \label{eq:main_result:periodic_stationary_state}
  \end{equation}
  Hence, this initial distribution $\vb{p}^\mathrm{ps}(0)$ is the
  eigenvector of $\vb{M}(\period)$ with eigenvalue one.  

  Using standard methods, as explained for example in
  Ref.~\cite{bara18a}, the generating function~\eqref{eq:setup:scgf}
  for the fluctuations of a general current
  observable~\eqref{eq:setup:time_add_curr} in the periodic stationary
  state is given by
  \begin{equation}
    \lambda_t(z)=\frac{1}{t}\ln\sum_{i,j}\mathcal{M}_{ij}(t,z)p_j^\mathrm{ps}(0),\quad
    \boldsymbol{\mathcal{M}}(t,z)\equiv \timeorderedexp{\int_0^t \dd{\tau}\boldsymbol{\mathcal{L}} (\tau,z)},
  \end{equation}
  with the tilted evolution operator $\boldsymbol{\mathcal{M}}(t,z)$
  and the tilted generator $\boldsymbol{\mathcal{L}}(\tau,z)$ with
  entries
  $\mathcal{L}_{ij}(\tau,z)\equiv
  L_{ij}(\tau)\exp(zd_{ji}(\tau))+\delta_{ij}\dot{a}_i(\tau)\,z$.
  In the long-time limit, the generating function follows as
  \begin{equation}
    \lambda(z)=[\ln\operatorname{eig}(\boldsymbol{\mathcal{M}}(\period,z))]/\period
  \end{equation}
  with $\operatorname{eig}(\boldsymbol{\mathcal{M}}(T,z))$ being the
  maximal eigenvalue of $\boldsymbol{\mathcal{M}}(T,z)$. For the
  illustration of the GTUR and the TUR for the two-level heat engine,
  we have calculated $\lambda(z)$ in a small region around $z=0$,
  yielding $D$ through numerical differentiation.

  \section{Limiting cases}
  \label{sec:appb}
  In the limit of fast driving, where $k_{ij}(t)\ll\omega$ for all
  transition rates at all times, the time-ordered exponential in
  \eqref{eq:main_result:time_ordered_exp} can be expanded as
  \begin{equation}
    \vb{M}(\period)=\vb{1}+\int_0^\period \dd{\tau}\matr{L} (\tau)+\mathcal{O}((k/\omega)^2),
  \end{equation}
  where $k$ stands generically for the scaling of all transition rates. 
  The eigenvalue
  equation~\eqref{eq:main_result:periodic_stationary_state} is then given
  in leading order by~\eqref{eq:main_result:p_eff_a}.  Then, the
  periodic stationary state $\vps(\tau)$ is in leading order
  time-independent and given by the first choice of the effective
  density $\vpeff$, i.e., $\vps(\tau)=\vpeff+\mathcal{O}(k/\omega).$
  Due to its time-independence, the leading order of
  $\vps(\tau)$ is also captured by the second
  choice for $\vpeff$~\eqref{eq:main_result:p_eff_b}. Consequently,
  the periodic stationary currents and the effective currents, while still being time-dependent, become equal
  in the leading zeroth order, i.e., $\js(\tau)=\jeff(\tau)+\mathcal{O}(k^2/\omega)$
  and thus $\sigma=\bound+\mathcal{O}(k^2/\omega)$ in Eq.~\eqref{eq:main_result:bound}.

  Another special case where the original form of the TUR is restored
  is the one where the transition rates become time-independent and
  correspond to a genuine non-equilibrium steady state, which leads to
  non-zero stationary currents
  $j_{ij}^\mathrm{s}=\js(\tau)=\jeff(\tau)$. Remarkably, this result
  goes beyond the classical statement of the TUR if the increments
  $d_{ij}(\tau)$ and $\dot{a}_i(\tau)$ are still periodically
  time-dependent. However, in the limiting case of time-independent
  transition rates that correspond to a non-driven system, the
  currents $\js(\tau)$ and $\jeff(\tau)$ both vanish and differ in
  leading order, such that $\bound$ does not approach $\sigma$.

  Finally, the limiting case $k_{ij}(t)\gg\omega$ with continuous
  protocols for $E_i(\tau)$ and $\beta(\tau)$ and in the absence of
  driving affinities presents the quasistatic limit, where the
  periodic stationary state
  \begin{equation}
    \ps_i(\tau)=p_i^\mathrm{eq}(\tau)+\mathcal{O}(\omega/k)\equiv\exp(-\beta(\tau)E_i(\tau))/Z(\tau)+\mathcal{O}(\omega/k)
  \end{equation}
  approaches an instantaneous equilibrium state normalized by the
  partition function~$Z(\tau)$. Since in the true equilibrium state
  corresponding to some point in time all currents would vanish, we
  obtain for the periodic stationary currents
  \begin{equation}
    \js(\tau)=\ps_i(\tau) k_{ij}(\tau)-\ps_j(\tau)k_{ji}(\tau)=\mathcal{O}(\omega),
  \end{equation}
  which is consistent with the condition
  $\dot{p}^\mathrm{s}_i(\tau)=-\sum_{j\neq i}\js(\tau)=\mathcal{O}(\omega)$.
  Using this scaling in Eq.~\eqref{eq:setup:definition_power} yields
  that the power scales like $\mathcal{O}(\omega)$.  The logarithm in
  Eq.~\eqref{eq:setup:definition_entropy} scales like
  $\mathcal{O}(\omega)$, such that $\sigma=\mathcal{O}(\omega^2/k)$.
  Note that in Figs.~\ref{fig:ill_and_appl:gtur_k0}
  and~\ref{fig:ill_and_appl:eta_heat_engine}, the two discontinuous
  jumps in $\beta(\tau)$ come with a finite production of entropy,
  leading to a dominant term in $\sigma$ that scales like
  $\mathcal{O}(\omega)$ and accordingly to an efficiency
  $\eta=P/(P+\sigma)$ that is finite and less than one.  The effective
  current~\eqref{eq:main_result:jeff} scales in leading order like
  $\mathcal{O}(k)$, leading to $\bound=\mathcal{O}(k^2/\omega)$ for
  the effective entropy production in
  Eq.~\eqref{eq:main_result:definition_of_bound}.  Fluctuations of the
  power, as quantified by $D_P$, tend to zero in the quasistatic limit,
  since the many jumps in any typical trajectory average the power
  in Eq.~\eqref{eq:setup:definition_power} to always the same
  value~\cite{spec04}. A quantitative analysis of the decay of the
  correlation function of the occupation observables $\occupation(\tau)$
  yields that $D_P=\mathcal{O}(\omega^2/k)$.

\end{appendix}


\section*{References}

\begin{thebibliography}{10}
\newcommand{\enquote}[1]{``#1''}
\newcommand{\doibase}[0]{http://doi.org/}

\bibitem{jarz11}
C.~{J}arzynski, \enquote{Equalities and inequalities: Irreversibility and the
  second law of thermodynamics at the nanoscale}, {\em Ann. Rev. Cond. Mat.
  Phys.\/} \href{\doibase10.1146/annurev-conmatphys-062910-140506}{{\bf 2},
  329} (2011).

\bibitem{seif12}
U.~Seifert, \enquote{Stochastic thermodynamics, fluctuation theorems, and
  molecular machines}, {\em Rep. Prog. Phys.\/}
  \href{\doibase10.1088/0034-4885/75/12/126001}{{\bf 75}, 126001} (2012).

\bibitem{bara15}
A.~C. Barato and U.~Seifert, \enquote{Thermodynamic uncertainty relation for
  biomolecular processes}, {\em Phys.\ Rev.\ Lett.\/}
  \href{\doibase10.1103/PhysRevLett.114.158101}{{\bf 114}, 158101} (2015).

\bibitem{ging16}
T.~R. Gingrich, J.~M. Horowitz, N.~Perunov, and J.~L. England,
  \enquote{Dissipation bounds all steady-state current fluctuations}, {\em
  Phys.\ Rev.\ Lett.\/} \href{\doibase10.1103/PhysRevLett.116.120601}{{\bf
  116}, 120601} (2016).

\bibitem{pole16}
M.~Polettini, A.~Lazarescu, and M.~Esposito, \enquote{Tightening the
  uncertainty principle for stochastic currents}, {\em Phys.\ Rev.\ E\/}
  \href{\doibase10.1103/PhysRevE.94.052104}{{\bf 94}, 052104} (2016).

\bibitem{nard17a}
C.~Nardini and H.~Touchette, \enquote{Process interpretation of current
  entropic bounds}, {\em Eur. Phys. J. B\/}
  \href{\doibase10.1140/epjb/e2017-80612-7}{{\bf 91}, 16} (2018).

\bibitem{dech17}
A.~Dechant and S.-i. Sasa, \enquote{Current fluctuations and transport
  efficiency for general {L}angevin systems}, {\em J. Stat. Mech. Theor.
  Exp.\/} \href{\doibase10.1088/1742-5468/aac91a}{p. 063209} (2018).

\bibitem{piet17}
P.~Pietzonka, F.~Ritort, and U.~Seifert, \enquote{Finite-time generalization of
  the thermodynamic uncertainty relation}, {\em Phys. Rev. E\/}
  \href{\doibase10.1103/PhysRevE.96.012101}{{\bf 96}, 012101} (2017).

\bibitem{horo17}
J.~M. Horowitz and T.~R. Gingrich, \enquote{Proof of the finite-time
  thermodynamic uncertainty relation for steady-state currents}, {\em Phys.
  Rev. E\/} \href{\doibase10.1103/PhysRevE.96.020103}{{\bf 96}, 020103} (2017).

\bibitem{garr17}
J.~P. Garrahan, \enquote{Simple bounds on fluctuations and uncertainty
  relations for first-passage times of counting observables}, {\em Phys.\ Rev.\
  E\/} \href{\doibase10.1103/PhysRevE.95.032134}{{\bf 95}, 032134} (2017).

\bibitem{ging17}
T.~R. Gingrich and J.~M. Horowitz, \enquote{Fundamental bounds on first passage
  time fluctuations for currents}, {\em Phys. Rev. Lett.\/}
  \href{\doibase10.1103/PhysRevLett.119.170601}{{\bf 119}, 170601} (2017).

\bibitem{shir17a}
N.~Shiraishi, \enquote{Finite-time thermodynamic uncertainty relation do not
  hold for discrete-time {M}arkov process}, {\em arXiv:1706.00892\/}  (2017).

\bibitem{proe17}
K.~Proesmans and C.~Van~den~Broeck, \enquote{Discrete-time thermodynamic
  uncertainty relation}, {\em EPL\/}
  \href{\doibase10.1209/0295-5075/119/20001}{{\bf 119}, 20001} (2017).

\bibitem{chiu18}
D.~Chiuchi\`u and S.~Pigolotti, \enquote{Mapping of uncertainty relations
  between continuous and discrete time}, {\em Phys. Rev. E\/}
  \href{\doibase10.1103/PhysRevE.97.032109}{{\bf 97}, 032109} (2018).

\bibitem{bran18}
K.~Brandner, T.~Hanazato, and K.~Saito, \enquote{Thermodynamic bounds on
  precision in ballistic multiterminal transport}, {\em Phys. Rev. Lett.\/}
  \href{\doibase10.1103/PhysRevLett.120.090601}{{\bf 120}, 090601} (2018).

\bibitem{maci18}
K.~Macieszczak, K.~Brandner, and J.~P. Garrahan, \enquote{Unified thermodynamic
  uncertainty relations in linear response}, {\em Phys. Rev. Lett.\/}
  \href{\doibase10.1103/PhysRevLett.121.130601}{{\bf 121}, 130601} (2018).

\bibitem{bara16}
A.~C. Barato and U.~Seifert, \enquote{Cost and precision of {B}rownian clocks},
  {\em Phys. Rev. X\/} \href{\doibase10.1103/PhysRevX.6.041053}{{\bf 6},
  041053} (2016).

\bibitem{bara17a}
A.~C. Barato and U.~Seifert, \enquote{Thermodynamic cost of external control},
  {\em New J. Phys.\/} \href{\doibase10.1088/1367-2630/aa77d0}{{\bf 19},
  073021} (2017).

\bibitem{bert18}
L.~Bertini, R.~Chetrite, A.~Faggionato, and D.~Gabrielli, \enquote{Level 2.5
  large deviations for continuous-time {M}arkov chains with time periodic
  rates}, {\em Ann. Henri Poincar{\'e}\/} \href{\doibase10.1007/s00023-018-0705-3}{{\bf 19}, 3197} (2018).

\bibitem{bara18a}
A.~C. Barato and R.~Chetrite, \enquote{Current fluctuations in periodically
  driven systems}, {\em J. Stat. Mech. Theor. Exp.\/}
  \href{\doibase10.1088/1742-5468/aabfc5}{p. 053207} (2018).

\bibitem{bara18b}
A.~C. Barato, R.~Chetrite, A.~Faggionato, and D.~Gabrielli, \enquote{Bounds on
  current fluctuations in periodically driven systems}, {\em New J. Phys.\/}
  \href{\doibase10.1088/1367-2630/aae512}{{\bf 20}, 103023} (2018).

\bibitem{piet15}
P.~Pietzonka, A.~C. Barato, and U.~Seifert, \enquote{Universal bounds on
  current fluctuations}, {\em Phys.\ Rev.\ E\/}
  \href{\doibase10.1103/PhysRevE.93.052145}{{\bf 93}, 052145} (2016).

\bibitem{piet16b}
P.~Pietzonka, A.~C. Barato, and U.~Seifert, \enquote{Universal bound on the
  efficiency of molecular motors}, {\em J.\ Stat.\ Mech.:\ Theor.\ Exp.\/}
  \href{\doibase10.1088/1742-5468/2016/12/124004}{p. 124004} (2016).

\bibitem{piet17a}
P.~Pietzonka and U.~Seifert, \enquote{Universal trade-off between power,
  efficiency and constancy in steady-state heat engines}, {\em Phys. Rev.
  Lett.\/} \href{\doibase10.1103/PhysRevLett.120.190602}{{\bf 120}, 190602}
  (2018).

\bibitem{schm08}
T.~Schmiedl and U.~Seifert, \enquote{Efficiency at maximum power: An
  analytically solvable model for stochastic heat engines}, {\em EPL\/}
  \href{\doibase10.1209/0295-5075/81/20003}{{\bf 81}, 20003} (2008).

\bibitem{blic12}
V.~Blickle and C.~Bechinger, \enquote{Realization of a micrometre-sized
  stochastic heat engine}, {\em Nature Phys.\/}
  \href{\doibase10.1038/nphys2163}{{\bf 8}, 143} (2012).

\bibitem{holu18}
V.~Holubec and A.~Ryabov, \enquote{Cycling tames power fluctuations near
  optimum efficiency}, {\em Phys. Rev. Lett.\/}
  \href{\doibase10.1103/PhysRevLett.121.120601}{{\bf 121}, 120601} (2018).

\bibitem{raz16a}
O.~Raz, Y.~Suba\c{s}\i{}, and C.~Jarzynski, \enquote{Mimicking nonequilibrium
  steady states with time-periodic driving}, {\em Phys. Rev. X\/}
  \href{\doibase10.1103/PhysRevX.6.021022}{{\bf 6}, 021022} (2016).

\bibitem{rots16}
G.~M. Rotskoff, \enquote{Mapping current fluctuations of stochastic pumps to
  nonequilibrium steady states}, {\em Phys.\ Rev.\ E\/}
  \href{\doibase10.1103/PhysRevE.95.030101}{{\bf 95}, 030101} (2017).

\bibitem{jarz97}
C.~{J}arzynski, \enquote{Nonequilibrium equality for free energy differences},
  {\em Phys.\ Rev.\ Lett.\/} \href{\doibase10.1103/PhysRevLett.78.2690}{{\bf
  78}, 2690} (1997).

\bibitem{shir16}
N.~Shiraishi, K.~Saito, and H.~Tasaki, \enquote{Universal trade-off relation
  between power and efficiency for heat engines}, {\em Phys.\ Rev.\ Lett.\/}
  \href{\doibase10.1103/PhysRevLett.117.190601}{{\bf 117}, 190601} (2016).

\bibitem{ging16a}
T.~R. Gingrich, G.~M. Rotskoff, and J.~M. Horowitz, \enquote{Inferring
  dissipation from current fluctuations}, {\em J. Phys. A: Math. Theor.\/}
  \href{\doibase10.1088/1751-8121/aa672f}{{\bf 50}, 184004} (2017).

\bibitem{spec04}
T.~Speck and U.~Seifert, \enquote{Distribution of work in isothermal
  nonequilibrium processes}, {\em Phys.\ Rev.\ E\/}
  \href{\doibase10.1103/PhysRevE.70.066112}{{\bf 70}, 066112} (2004).

\end{thebibliography}
\end{document}